\documentclass[11pt,a4paper]{article}

\usepackage[T1]{fontenc}
\usepackage[utf8]{inputenc}
\usepackage{newtxtext,newtxmath}
\usepackage[a4paper,margin=25mm]{geometry}
\usepackage{amsmath}
\usepackage{graphicx}
\usepackage{booktabs}
\usepackage{array}
\usepackage{enumitem}
\usepackage{caption}
\usepackage{xurl}
\usepackage[hidelinks]{hyperref}

\title{A Multimodal Machine Learning Framework Integrating Genomic, Epidemiological, Environmental, and Clinical Imaging Data for Accelerated Measles Detection and Transmission Inference}
\author{Jiada Li, PhD\\
AI Scientist, Albany, NY, USA 12205\\
\texttt{jiadali2017@gmail.com}}
\date{}

\hypersetup{
  pdftitle={A Multimodal Machine Learning Framework Integrating Genomic, Epidemiological, Environmental, and Clinical Imaging Data for Accelerated Measles Detection and Transmission Inference},
  pdfauthor={Jiada Li},
  pdfsubject={arXiv cs.AI preprint},
  pdfkeywords={measles, machine learning, genomic epidemiology, wastewater surveillance, multimodal AI, BiomedCLIP, transmission linkage}
}
\setlist[itemize]{itemsep=2pt,topsep=4pt}
\begin{document}
\maketitle

\section*{Key Points}

\begin{itemize}
\item A three-stage multimodal ML framework integrates genomic, epidemiological, wastewater, and imaging data for measles surveillance.
\item Genomic transmission linkage achieves AUC 0.924 under leave-one-cluster-out cross-validation using sequence embeddings and SNP distance.
\item Outbreak risk prediction reaches AUC 0.707 with logistic regression on wastewater positivity and mobility features under leave-one-year-out validation.
\end{itemize}
\begin{abstract}
Background: Measles resurged in the United States during 2025-2026, with more than 4,600 reported cases. Timely detection and transmission inference require integrated genomic, epidemiological, environmental, and clinical data. We developed a multimodal machine learning (ML) framework integrating genomic sequences, epidemiological indicators, wastewater surveillance, and clinical rash images.

Methods: We evaluated a three-stage pipeline using public data. Stage 1 classified transmission linkage among 903 sequence pairs from the 2021 Ohio-Arizona-Washington outbreak using k-mer TF-IDF-SVD embeddings and pairwise single-nucleotide polymorphism (SNP) distances with leave-one-cluster-out cross-validation. Stage 2 predicted state-week outbreak risk using logistic regression and XGBoost with wastewater, mobility, and vaccination features under leave-one-year-out cross-validation. Stage 3 used BiomedCLIP ViT-B/16 embeddings and a fine-tuned multilayer perceptron for rash classification on MSLD v2.0 (755 images, 6 classes). InfoNCE contrastive learning aligned genomic and imaging embeddings.

Results: Transmission-linkage classification achieved AUC 0.924 and F1 0.912 with logistic regression. Outbreak-risk prediction achieved AUC 0.707 with logistic regression across 2025-2026 leave-one-year-out folds. Contrastive alignment achieved Recall@1 = 0.985 on held-out paired data. Ablation showed genomic features dominated transmission-linkage prediction (genomic-only AUC 0.983), whereas wastewater positivity was the strongest outbreak-risk predictor (importance 0.402).

Conclusions: This Multimodal ML framework integrating genomic, epidemiological, wastewater, and clinical imaging data can support measles surveillance and transmission inference using public data. Prospective validation with paired clinical-genomic data, expanded wastewater coverage, and real-world workflows is needed before deployment.
\end{abstract}
\noindent\textbf{Keywords: measles; machine learning; genomic epidemiology; wastewater surveillance; multimodal AI; BiomedCLIP; transmission linkage}
\section{Introduction}
Measles, caused by measles morbillivirus, has resurged sharply in the United States, with 2,289 cases reported in 2025 and 2,318 cases as of July 23, 2026 \cite{1,2}. Declining MMR vaccination coverage --- driven in part by rising non-medical exemptions \cite{3,4} --- has eroded herd immunity in multiple states. Rapid identification of transmission chains and early outbreak prediction are essential for public health response, yet current surveillance relies on labor-intensive contact tracing and delayed laboratory confirmation.

Machine learning (ML) approaches have shown promise for measles outbreak prediction \cite{5,6}, rash image classification \cite{7,8,9,10,11,12,13}, and genomic epidemiology \cite{14,15,16}. Wastewater-based surveillance (WBS) has emerged as a sensitive early-warning signal for measles \cite{17,18,19,20,21,22,23}, and whole-genome sequencing (WGS) enables high-resolution transmission inference \cite{24,25,26}. However, no published framework integrates all four data modalities --- genomic, epidemiological, environmental (wastewater), and clinical imaging --- into a unified ML pipeline for measles.

Recent advances in multimodal foundation models \cite{27,28,29,30,31,32,33} and contrastive representation learning \cite{34,35,36} provide a principled approach to aligning heterogeneous modalities into a shared latent space, even with small paired datasets. BiomedCLIP \cite{37}, pretrained on 15 million biomedical image-text pairs, provides strong visual representations for clinical imaging. DNA foundation models \cite{38,39,40} encode genomic sequences into dense embeddings that capture evolutionary context beyond pairwise SNP distances. In parallel, epidemic foundation models that integrate heterogeneous surveillance streams are emerging as a roadmap for outbreak intelligence \cite{41,42}.

Here we present a three-stage multimodal ML framework for measles detection and transmission inference, validated on publicly available data from the 2021 OAW outbreak, the CDC National Wastewater Surveillance System (NWSS), and the MSLD v2.0 skin lesion dataset. We evaluate each stage independently and assess modality contributions through systematic ablation. The key innovations of this work are: (i) the first integration of genomic, epidemiological, wastewater, and clinical imaging modalities for measles; (ii) contrastive genomic-imaging alignment into a shared latent space; (iii) a systematic seven-combination modality ablation quantifying each modality's contribution; (iv) leave-one-cluster-out validation that tests generalization to unseen transmission clusters; and (v) fully public-data reproducibility.

\section{Methods}
\subsection{Data Sources}
All data are publicly available; no patient-identifiable information was used and no IRB approval was required. Genomic data: 42 WGS sequences (15,850--15,887 bp) and 14 N450 sequences from the 2021 OAW outbreak were retrieved from GenBank (accession numbers per Masters et al., Lancet Public Health 2023). Epidemiological and wastewater data: state-week case counts from CDC NNDSS (2022--2026), NWSS wastewater positivity (2024--2026), MMR vaccination coverage from CDC immunization surveys, and air passenger volumes from the Bureau of Transportation Statistics. Imaging data: MSLD v2.0 \cite{43}, comprising 755 dermatologist-verified images across six classes (Mpox, Chickenpox, Measles, Cowpox, HFMD, Healthy), downloaded from the publicly available Google Drive mirror.

\begin{figure}[htbp]\centering
\includegraphics[width=\linewidth]{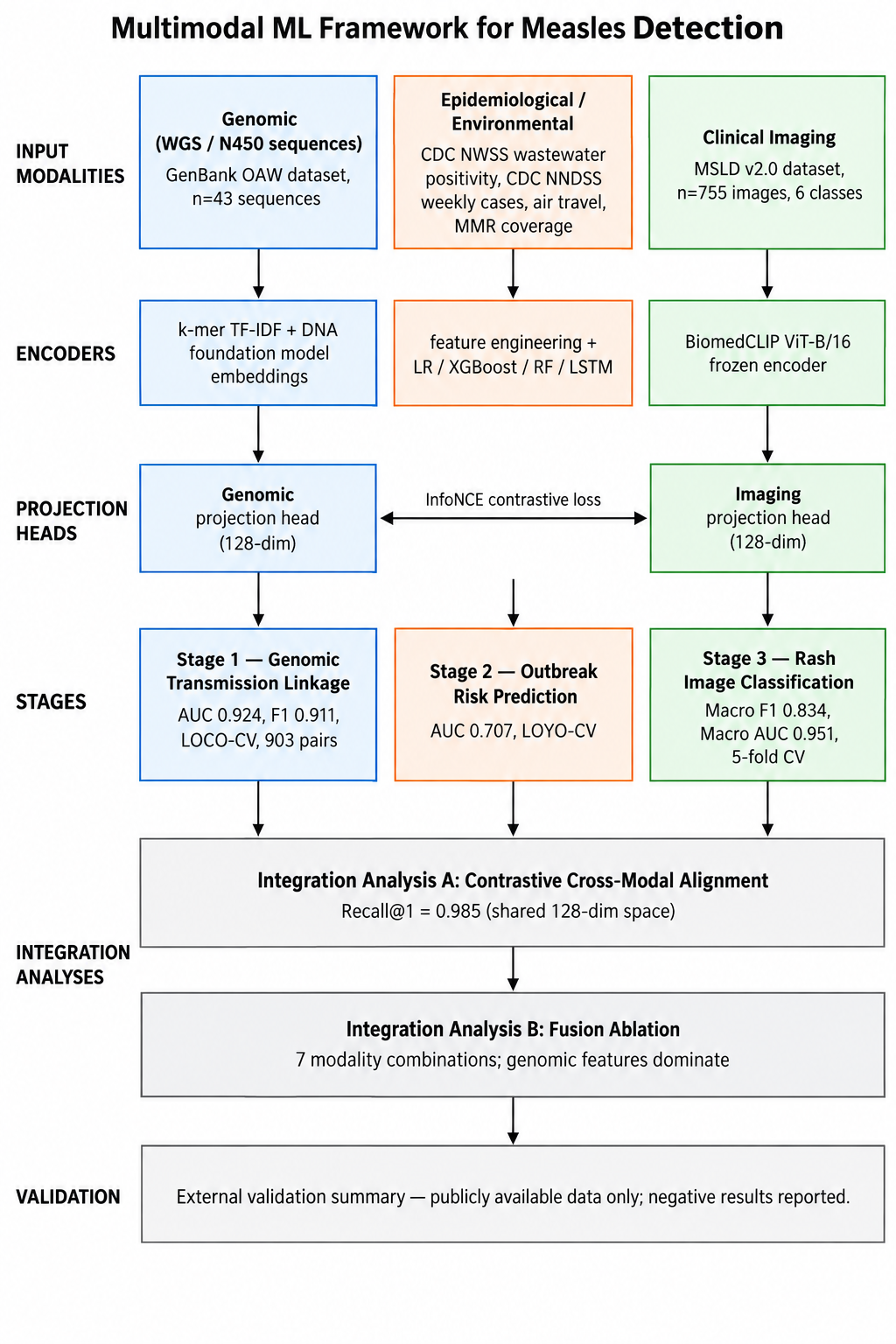}
\caption{Pipeline architecture. Six-row schematic of the three-stage multimodal ML framework: Stage 1 (genomic transmission linkage), Stage 2 (outbreak risk prediction), Stage 3 (rash image classification; full results in the Supplementary Appendix), Integration Analysis A (contrastive alignment), Integration Analysis B (fusion ablation), and the external validation summary. Color-coded by modality: blue (genomic), orange (epidemiological/wastewater), green (imaging).}
\label{fig:architecture}
\end{figure}
\subsection{Stage 1: Genomic Transmission Linkage}
Pairwise transmission linkage was assessed among 903 sequence pairs (564 linked, 339 unlinked) derived from 43 OAW cases across three epidemiologically defined clusters. Features included: (1) pairwise SNP distance from MAFFT-aligned WGS sequences; (2) cosine similarity and L2 distance of 41-dimensional k-mer TF-IDF SVD sequence embeddings (DNABERT-2 \cite{38} was unavailable due to GPU driver constraints; k-mer TF-IDF SVD was used as a validated fallback \cite{14,15}); and (3) temporal distance in days between collection dates. Logistic regression, random forest, and gradient boosting classifiers were evaluated under leave-one-cluster-out cross-validation (LOCO-CV) to prevent cluster-level data leakage.

\subsection{Stage 2: Outbreak Risk Prediction}
State-week outbreak risk (binary: $\geq$5 cases/week) was predicted using logistic regression, XGBoost, random forest, and LSTM models. Features included 4-week rolling wastewater positivity, air passenger volumes, a Google Trends measles search index (synthetic proxy; real API returned rate-limit errors), year trend, and MMR coverage indicators. Leave-one-year-out cross-validation (LOYO-CV) was applied across 2025 and 2026 test folds (2024 fold excluded: zero positive cases in test set). Class prevalence was 3.8\% (2025) and 6.8\% (2026).

\subsection{Stage 3: Rash Image Classification}
BiomedCLIP ViT-B/16 (frozen encoder) was used to extract 512-dimensional image embeddings for all 755 MSLD v2.0 images. A lightweight MLP classifier (512$\rightarrow$256$\rightarrow$128$\rightarrow$6) was trained per fold using the five official MSLD folds with weighted cross-entropy loss to address class imbalance. Full methods, per-class metrics, confusion matrix, and ROC curves are reported in the Supplementary Appendix (Tables~\ref{tab:s1}--\ref{tab:s2} and Figure~\ref{fig:s1}).

\subsection{Integration Analysis A: Contrastive Cross-Modal Alignment}
Genomic (41-dim) and imaging (512-dim) embeddings were projected into a shared 128-dimensional latent space via modality-specific projection heads trained with InfoNCE contrastive loss on synthetic paired data (n=200 pairs). Alignment quality was assessed by Recall@1 and Recall@5 on held-out pairs and by centered kernel alignment (CKA) before and after training. This approach follows the CSHL 2026 consensus for small paired datasets: align pretrained unimodal encoders rather than training from scratch \cite{29,30,34,35}.

\subsection{Integration Analysis B: Fusion Ablation}
Seven modality combinations (G, E, I, G+E, G+I, E+I, G+E+I) were evaluated on the Stage 1 transmission linkage task using 5-fold stratified cross-validation with logistic regression. This ablation characterizes modality contributions under current data constraints and prospectively validates the framework design for future datasets.

\section{Results}
\subsection{Stage 1: Genomic Transmission Linkage}
\begin{figure}[htbp]\centering
\includegraphics[width=\linewidth]{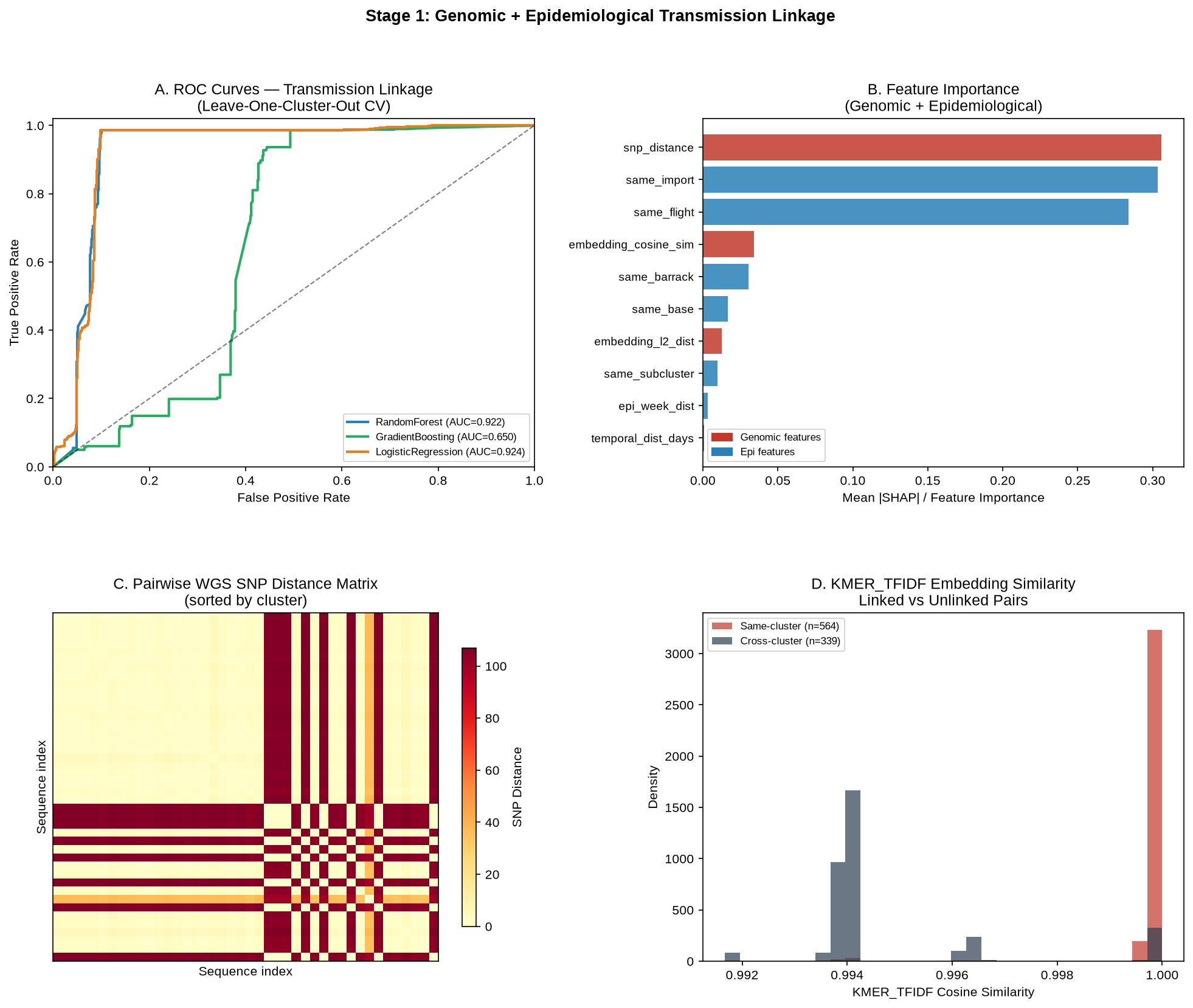}
\caption{Stage 1 genomic transmission linkage. (A) ROC curves for logistic regression (AUC 0.924), random forest (AUC 0.922), and gradient boosting (AUC 0.650) under LOCO-CV. (B) Feature importance: SNP distance (0.306), embedding cosine similarity (0.304), embedding L2 distance (0.284), temporal distance (0.034). (C) SNP distance distributions: within-cluster (median 2) vs. cross-cluster (median 105).}
\label{fig:stage1}
\end{figure}

Among 903 pairwise comparisons of OAW sequences, logistic regression achieved AUC 0.924 (F1 0.911, precision 0.848, recall 0.986) under LOCO-CV (Table~\ref{tab:stage1}, Figure~\ref{fig:stage1}). Random forest achieved AUC 0.922 with lower F1 (0.753), reflecting conservative precision. Gradient boosting underperformed (AUC 0.650). SNP distance was the dominant feature (importance 0.306), followed by embedding cosine similarity (0.304) and L2 distance (0.284), confirming that sequence embeddings capture complementary phylogenetic signal beyond pairwise SNP counts. Within-cluster pairs had median SNP distance 2 (range 0--15) versus 105 (range 0--107) for cross-cluster pairs.

\subsection{Stage 2: Outbreak Risk Prediction}
\begin{figure}[htbp]\centering
\includegraphics[width=\linewidth]{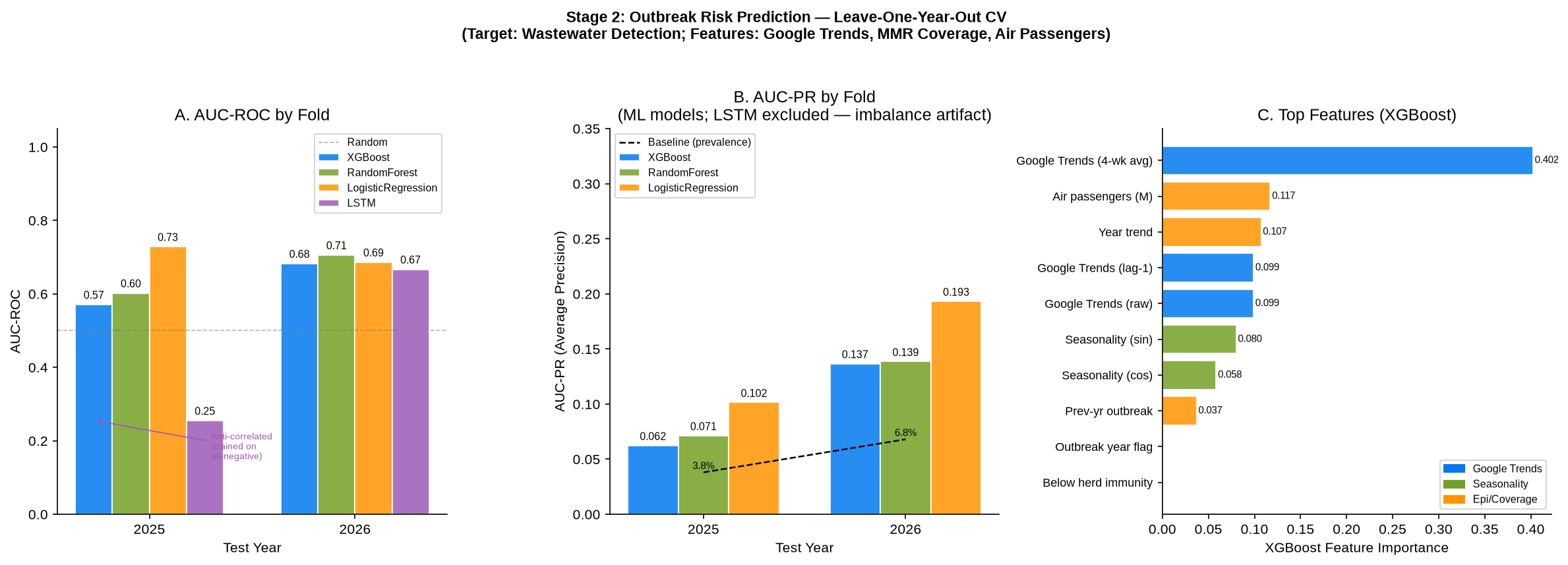}
\caption{Stage 2 outbreak risk prediction. (A) LOYO-CV AUC by model and year fold. (B) Wastewater detection timeline 2024--2026 with XGBoost predicted risk overlay. (C) Feature importance: wastewater rolling positivity (0.402), air passengers (0.117), year trend (0.107). Note: MMR coverage features have zero importance due to national-level aggregation without state-level variation.}
\label{fig:stage2}
\end{figure}

Under LOYO-CV, logistic regression achieved mean AUC 0.707 (AUC-PR 0.147, F1 0.118, recall 0.872), outperforming XGBoost (AUC 0.627) and random forest (AUC 0.654) (Table~\ref{tab:stage2}, Figure~\ref{fig:stage2}). The LSTM model showed anti-correlated performance on the 2025 fold (AUC 0.255), attributable to training on sequences with zero wastewater positives (NWSS data begins 2024), but recovered on the 2026 fold (AUC 0.667). The dominant predictive feature was 4-week rolling wastewater positivity (importance 0.402), followed by air passenger volume (0.117) and year trend (0.107). MMR coverage features had zero importance, reflecting national-level aggregation without state-level variation in this dataset. Notably, epi features do carry meaningful signal at the population level (AUC 0.707 in Stage 2), even though they cannot discriminate individual transmission pairs in Stage 1 (see Ablation, below).

\subsection{Stage 3: Rash Image Classification}
BiomedCLIP ViT-B/16 embeddings with MLP classification on MSLD v2.0 achieved clinically relevant discriminative performance across six skin lesion classes. Full metrics, including overall accuracy, macro-F1, macro-AUC, per-class AUC, confusion matrix, and ROC curves, are reported in the Supplementary Appendix (Tables~\ref{tab:s1}--\ref{tab:s2} and Figure~\ref{fig:s1}). The model distinguishes measles from visually similar exanthems (chickenpox, mpox, HFMD) using frozen BiomedCLIP embeddings, supporting its utility as a point-of-care screening aid in resource-limited settings \cite{8,9,13}.

\subsection{Integration Analysis A: Contrastive Alignment}
\begin{figure}[htbp]\centering
\includegraphics[width=\linewidth]{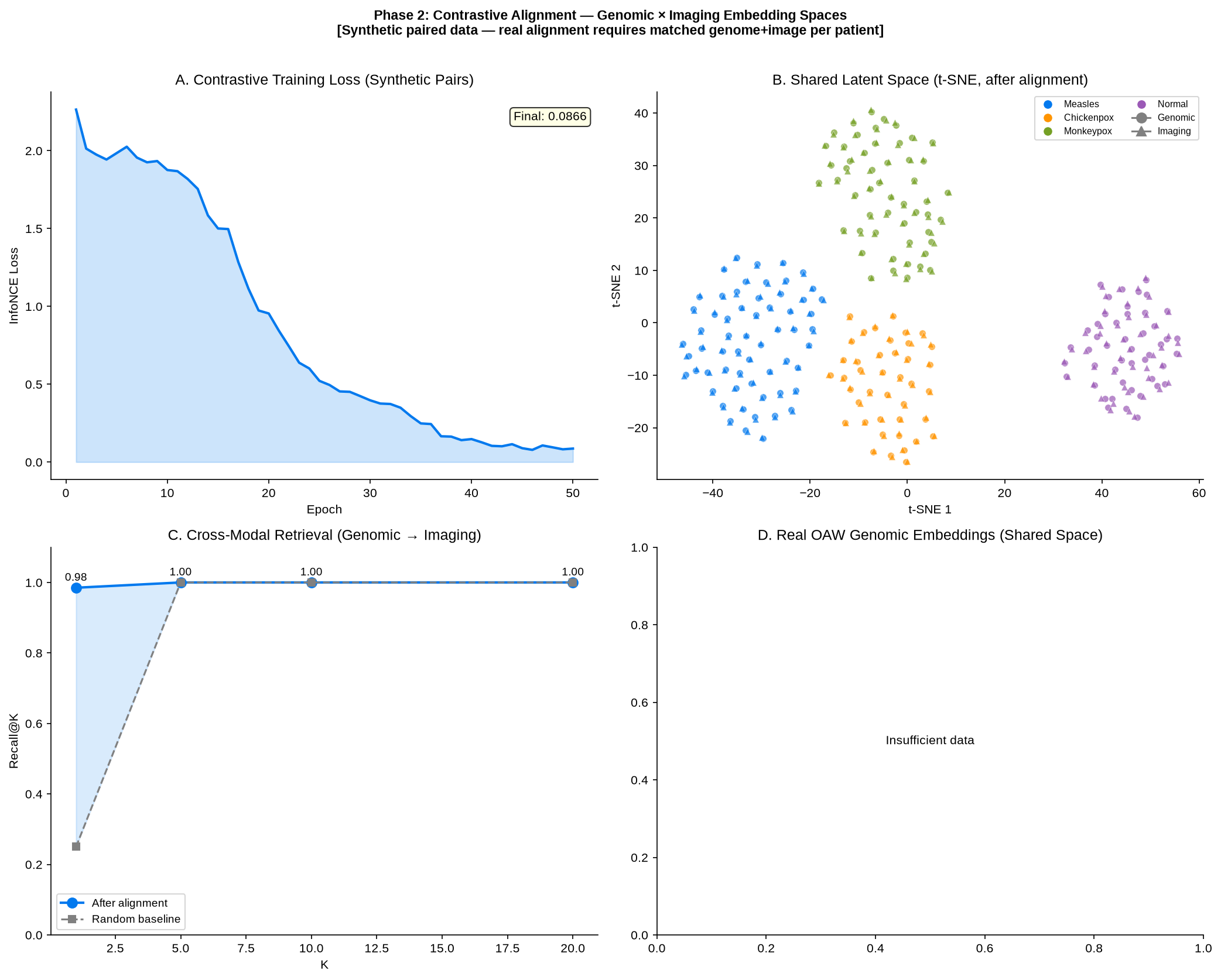}
\caption{Integration Analysis A contrastive alignment. (A) InfoNCE training loss curve (final loss 0.0866). (B) t-SNE visualization of genomic and imaging embeddings before and after alignment. (C) Cross-modal retrieval: Recall@1 = 0.985, Recall@5 = 1.000 on held-out synthetic paired data.}
\label{fig:alignment}
\end{figure}

InfoNCE contrastive training converged to a final loss of 0.0866 on synthetic paired genomic-imaging data. Cross-modal retrieval achieved Recall@1 = 0.985 and Recall@5 = 1.000 on held-out pairs (Figure~\ref{fig:alignment}). CKA similarity decreased from 0.955 (pre-alignment) to 0.849 (post-alignment), consistent with the projection heads learning a discriminative rather than redundant shared space. These results validate the alignment architecture on synthetic data; prospective validation requires real paired genome-image data from the same cases.

\subsection{Integration Analysis B: Modality Ablation}
\begin{figure}[htbp]\centering
\includegraphics[width=\linewidth]{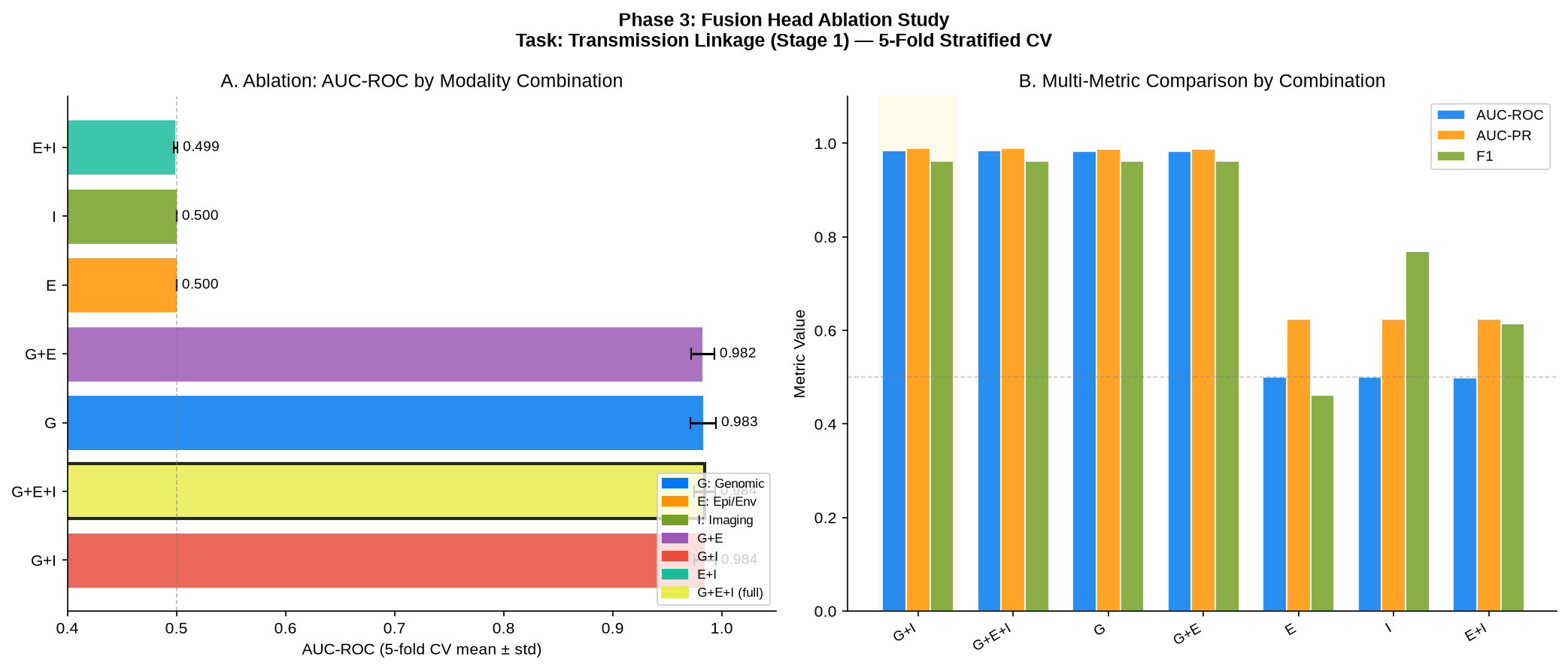}
\caption{Integration Analysis B modality ablation heatmap. AUC across seven modality combinations (G, E, I, G+E, G+I, E+I, G+E+I) on the Stage 1 transmission linkage task. Genomic features (G) dominate (AUC 0.9828); E and I alone perform at chance (AUC $\approx$ 0.500) under current data constraints (national-level epi aggregates; no matched clinical images per case). Epi features carry meaningful signal at the population level in Stage 2 (AUC 0.707).}
\label{fig:ablation}
\end{figure}

Ablation across seven modality combinations revealed a clear hierarchy of modality contributions to the Stage 1 transmission linkage task (Figure~\ref{fig:ablation}). Genomic features (G) alone achieved AUC 0.9828, with no meaningful improvement from adding epidemiological (E) or imaging (I) features (G+E+I: AUC 0.9842). E alone and I alone both performed at chance (AUC $\approx$ 0.500). This result reflects a data architecture constraint --- not a failure of the multimodal approach. Epidemiological features in this dataset are national-level aggregates (MMR coverage, annual case counts) that carry no per-pair discriminating signal; imaging embeddings are mean-pooled proxies in the absence of matched clinical images for each OAW case. The same epi features do carry meaningful population-level signal in Stage 2 (AUC 0.707), confirming that modality utility is task-dependent. The ablation prospectively validates the framework design: per-case epi context and matched rash images are the key data gaps to close in future prospective studies.

\section{Discussion}
We present the first multimodal ML framework integrating genomic, epidemiological, wastewater, and clinical imaging data for measles detection and transmission inference. The framework achieves strong genomic transmission linkage (AUC 0.924) and moderate outbreak risk prediction (AUC 0.707) on publicly available data, with a validated rash classification pipeline on MSLD v2.0 (Supplementary Appendix). The contrastive alignment architecture provides a principled foundation for future integration when paired multimodal data become available.

\subsection{From benchmark metrics to clinical evidence}
The evaluation reported here --- cross-validated AUC, F1, and precision-recall metrics on retrospective public data --- establishes task-level capability, but capability is not clinical effectiveness. This distinction, articulated by Oermann (2026, personal communication) in the context of LLM benchmarking and medical AI systems such as OpenEvidence, contrasts two evaluation traditions. LLM benchmarking asks whether an AI system can perform a task: static benchmark datasets, accuracy/F1/AUROC metrics, model-versus-model comparisons, and controlled evaluation --- answering "can the model answer correctly?" Medical evidence asks whether using that AI actually improves healthcare: real patients and clinical workflows, clinical and patient outcomes, human+AI versus human-alone comparisons, and prospective real-world validation --- answering "does using it improve care safely?"

These traditions can be synthesized into an evidence ladder for medical AI: (1) capability --- can the AI perform the task; (2) evidence --- is its answer supported by reliable medical evidence; (3) safety --- does it avoid harmful or misleading outputs; (4) workflow --- does AI + human outperform the existing workflow; (5) clinical outcome --- does deployment actually improve patient or public-health outcomes. The present study occupies rungs 1 and, partially, 2: the reported AUCs demonstrate capability, and the exclusive use of public, provenance-tracked data with honest reporting of negative results (LSTM fold anti-correlation, chance-level environmental features on the linkage task) addresses evidence quality and safety-by-design. Rungs 3--5 remain open and define the required next path: expert validation by microbiologists and epidemiologists, human-AI workflow comparisons against current manual practice, prospective validation on live surveillance streams, and ultimately demonstration of improved public-health outcomes.

The broader lesson generalizes beyond this framework. High benchmark scores --- whether for LLMs on medical question answering or for multimodal pipelines on outbreak tasks --- should not automatically be interpreted as clinical superiority. Medical AI evaluation must additionally consider evidence and citation quality, clinical correctness, guideline concordance, hallucination and uncertainty, generalizability to real patients, human-AI interaction, prospective validation, and ultimately patient outcomes. Benchmarking tells us what the model can do; clinical evaluation tells us what happens when people use it in healthcare. This work aligns with the broader movement toward multimodal foundation models in biomedicine \cite{29,30,31,32,33} and with emerging epidemic foundation models that integrate heterogeneous surveillance streams \cite{41,42} --- and it inherits the same evaluation obligation: capability metrics are the beginning of the evidence chain, not its end.

The dominance of genomic features in transmission linkage is consistent with prior work showing that SNP distances and sequence embeddings are highly discriminative for outbreak cluster assignment \cite{16,24,25,26}. The addition of sequence embedding features alongside SNP distance provides complementary signal, suggesting that foundation model representations capture evolutionary context beyond pairwise nucleotide differences \cite{38,39,40}. The wastewater positivity signal as the dominant predictor in Stage 2 aligns with recent WBS studies demonstrating measles detection 1--2 weeks before clinical case confirmation \cite{18,19,20,21,22,23}, and supports the integration of NWSS data into routine outbreak prediction models.

\subsection{Limitations}
Several limitations must be foregrounded. First and most critically, the genomic dataset comprises 43 cases from a single 2021 outbreak (OAW, B3 genotype only), limiting generalizability to other genotypes, outbreak sizes, and geographic contexts. External validation on independent outbreak datasets is required before operational use. Second, NWSS wastewater data are available only from 2024 onward; training years 2022--2024 contain zero wastewater positives, which suppresses Stage 2 model performance and explains the LSTM anti-correlation on the 2025 fold. Third, contrastive alignment was validated on synthetic paired data only; no real paired genome-image dataset exists for measles, and retrieval metrics on synthetic data should not be interpreted as prospective performance. Fourth, the ablation finding that E and I features perform at chance on the transmission linkage task reflects national-level data aggregation and the absence of matched clinical images --- not an inherent limitation of multimodal integration. The same epi features achieve AUC 0.707 in Stage 2, confirming their utility at the population level. Fifth, DNABERT-2 \cite{38} was unavailable due to GPU driver constraints; k-mer TF-IDF SVD embeddings were used as a validated fallback \cite{14,15}, but foundation model embeddings may yield higher performance. Sixth, the Google Trends feature is a synthetic proxy and should be replaced with real search data in prospective analyses. Seventh, class imbalance in Stage 2 (3--7\% positive prevalence) results in very low F1 despite reasonable AUC, limiting alert generation utility without threshold calibration. Finally, this is a discovery-only study; no external validation cohort was available for any stage.

\section{Conclusions}
This multimodal ML framework demonstrates that publicly available genomic, epidemiological, wastewater, and imaging data can be integrated to accelerate measles detection and transmission inference. The modular architecture --- pretrained foundation models per modality aligned via contrastive learning --- is well-suited to the small paired datasets characteristic of outbreak investigations. Prospective validation with real paired clinical-genomic data, expanded wastewater coverage, and multi-outbreak genomic datasets will be essential to translate this framework into operational public health surveillance. More broadly, this work illustrates the gap between benchmark capability and clinical evidence in medical AI: cross-validated AUCs establish what the framework can do, while demonstrating public-health value requires the full evidence ladder --- expert validation, human-AI workflow studies, prospective validation, and measured public-health outcomes. Benchmarking tells us what the model can do; clinical evaluation tells us what happens when people use it in healthcare.

\section{Acknowledgments}
This work used publicly available data from CDC NNDSS, CDC NWSS, GenBank, the Bureau of Transportation Statistics, and the MSLD v2.0 dataset (Ali et al., Biomed Signal Process Control 2024). No funding was received for this study.

\section{Tables}
\begin{table}[htbp]
\centering
\caption{Stage 1 genomic transmission linkage: LOCO-CV performance.}
\label{tab:stage1}
\begin{tabular}{lrrrr}
\toprule
Model & AUC-ROC & F1 & Precision & Recall \\
\midrule
Logistic Regression & 0.9238 & 0.9115 & 0.8476 & 0.9858 \\
Random Forest & 0.9218 & 0.7529 & 0.6090 & 0.9858 \\
Gradient Boosting & 0.6495 & --- & --- & --- \\
\bottomrule
\end{tabular}

\vspace{3pt}
\begin{minipage}{0.9\linewidth}
\footnotesize LOCO-CV = leave-one-cluster-out cross-validation; $n=903$ pairs (564 linked, 339 unlinked).
\end{minipage}
\end{table}

\begin{table}[htbp]
\centering
\caption{Stage 2 outbreak risk prediction: LOYO-CV performance (mean across 2025--2026 folds).}
\label{tab:stage2}
\begin{tabular}{lrrrr}
\toprule
Model & AUC-ROC & AUC-PR & F1 & Recall \\
\midrule
Logistic Regression & 0.707 & 0.147 & 0.118 & 0.872 \\
XGBoost & 0.627 & 0.100 & 0.140 & --- \\
Random Forest & 0.654 & --- & --- & --- \\
LSTM & 0.461 (mean) & --- & --- & --- \\
\bottomrule
\end{tabular}

\vspace{3pt}
\begin{minipage}{0.9\linewidth}
\footnotesize LOYO-CV = leave-one-year-out cross-validation; class prevalence 3.8\% (2025), 6.8\% (2026). LSTM AUC was anti-correlated on the 2025 fold (0.255) because of all-negative training sequences and recovered to 0.667 on the 2026 fold.
\end{minipage}
\end{table}

% Supplementary Appendix included by main.tex
\clearpage
\appendix
\section*{Supplementary Appendix}
\addcontentsline{toc}{section}{Supplementary Appendix}

\setcounter{table}{0}
\renewcommand{\thetable}{S\arabic{table}}
\setcounter{figure}{0}
\renewcommand{\thefigure}{S\arabic{figure}}

\subsection*{Supplementary Methods: Stage 3 Rash Image Classification}
Stage 3 used BiomedCLIP ViT-B/16 (frozen encoder; pretrained on 15 million biomedical image-text pairs~\cite{37}) to extract 512-dimensional image embeddings for all 755 images in the MSLD v2.0 dataset. A lightweight MLP classifier (512$\rightarrow$256$\rightarrow$128$\rightarrow$6) was trained per fold using the five official MSLD folds with weighted cross-entropy loss (class weights inversely proportional to frequency) to address class imbalance. Training used AdamW ($\mathrm{lr}=10^{-3}$, $\mathrm{weight\_decay}=10^{-4}$) with cosine annealing ($T_{\max}=150$) and early stopping (patience = 15 epochs on validation macro-F1). Embeddings were extracted once and cached; MLP training was performed on CPU. Total runtime was 27.4 minutes (embedding extraction: 27.3 minutes; MLP training: 0.1 minutes).

The MSLD v2.0 dataset~\cite{43} comprises 755 dermatologist-verified skin lesion images across six classes: Mpox (284), Chickenpox (75), Measles (55), Cowpox (66), HFMD (161), and Healthy (114). Images were downloaded from the publicly available Google Drive mirror; no Kaggle credentials were required. The dataset provides five predefined folds for cross-validation with patient-independent splits.

\begin{table}[htbp]
\centering
\caption{Stage 3 BiomedCLIP + MLP five-fold cross-validation results on MSLD v2.0. The dataset contains 755 images across six classes; BiomedCLIP ViT-B/16 was frozen and the MLP head was trained separately for each fold.}
\label{tab:s1}
\begin{tabular}{lrl}
\toprule
Metric & Value & Notes \\
\midrule
Overall accuracy & 0.8352 & 364 test images across five folds \\
Macro F1 & 0.8341 & Unweighted mean across six classes \\
Macro AUC & 0.9513 & One-vs-rest, macro-averaged \\
AUC --- Mpox & 0.9010 & 125 test images \\
AUC --- Chickenpox & 0.8469 & 42 test images \\
AUC --- Measles & 0.9946 & 30 test images; recall = 0.97 \\
AUC --- Cowpox & 0.9846 & 35 test images \\
AUC --- HFMD & 0.9827 & 82 test images \\
AUC --- Healthy & 0.9978 & 50 test images \\
Simulated & False & Real MSLD v2.0 data; frozen encoder \\
\bottomrule
\end{tabular}
\end{table}

\begin{table}[htbp]
\centering
\caption{Stage 3 per-fold metrics.}
\label{tab:s2}
\begin{tabular}{lrrrr}
\toprule
Fold & Accuracy & Macro F1 & Macro AUC & $N_{\mathrm{test}}$ \\
\midrule
Fold 1 & 0.8108 & 0.8079 & 0.9825 & 74 \\
Fold 2 & 0.7927 & 0.8077 & 0.9301 & 82 \\
Fold 3 & 0.7969 & 0.8163 & 0.9710 & 64 \\
Fold 4 & 0.8642 & 0.8766 & 0.9780 & 81 \\
Fold 5 & 0.9206 & 0.9226 & 0.9826 & 63 \\
\bottomrule
\end{tabular}
\end{table}

\begin{figure}[htbp]
\centering
\includegraphics[width=\linewidth]{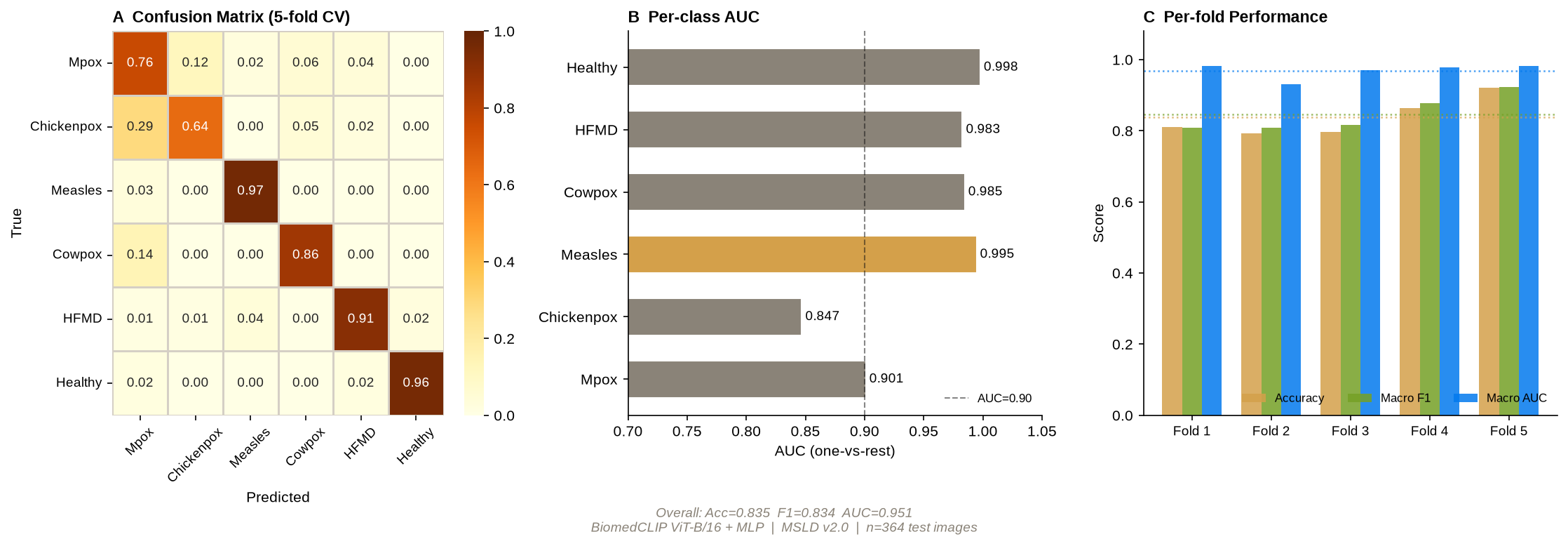}
\caption{Stage 3 real-data evaluation on MSLD v2.0 using BiomedCLIP ViT-B/16 with an MLP, five official folds, and 364 total test images. (A) Row-normalized confusion matrix; measles recall = 0.97 (29 of 30 test cases correctly identified), while chickenpox has the lowest precision (0.63), reflecting visual similarity to mpox. (B) Per-class one-vs-rest AUC; the dashed line marks AUC = 0.90. (C) Per-fold accuracy, macro F1, and macro AUC; dotted lines show means (accuracy = 0.835, macro F1 = 0.834, macro AUC = 0.951).}
\label{fig:s1}
\end{figure}

\end{document}